\documentclass{article}  
\usepackage{bigsky2007}
\usepackage{graphicx}
\usepackage{subfigure}
\usepackage{amssymb}
\frompage{000} \topage{000}                                              %|
\def\rightmark{Topological reconstruction of open charm mesons }
\def\leftmark{A. Mischke et al.}

\newcommand {\snn}      {\sqrt{s_{_{\rm NN}}}}

\newcommand {\pt}       {p_{\rm T}}

\newcommand {\deta}     {\Delta\eta}
\newcommand {\dphi}     {\Delta\phi}

\title{Topological reconstruction of open charm mesons using electron tagging} 
\authors{
{Andr\'e Mischke$^1$ for the STAR Collaboration
}\\[2.812mm]
{\normalsize
\hspace*{-8pt}$^1$ Institute for Subatomic Physics, Utrecht University, 
Princetonplein 5, 3584 CC Utrecht, The Netherlands\\
}}

\abstract{
We present first results on the topological reconstruction of open charm mesons in 
p+p collisions at $\snn$~=~200 GeV using electron tagging.
The analysis makes use of the full acceptance of the STAR electromagnetic calorimeter 
during Run VI data taking. 
A clear D$^0$ signal is obtained with a remarkable signal-to-background ratio of about 
1/7 and a signal significance of about 4. 
The azimuthal correlation distribution of the subleading electrons associated with 
open charm mesons exhibits a two-peak structure. 
We found first indications for prompt charm meson pair production.
This correlation technique allows detailed energy loss measurements of open charm
mesons in heavy-ion collisions.
}

 
\begin{document}
 
\maketitle
\setcounter{page}{1}

%-----------------------------------------------------------------------------------------
\section{Introduction}\label{intro}
Measurements at the Relativistic Heavy-Ion Collider (RHIC) at Brookhaven 
National Laboratory have revealed a strong modification of the jet structure
inside the created medium. Theoretical models, that attribute the jet 
attenuation to the energy loss of partons in the medium, have successfully 
described the present results for light-quark hadrons~\cite{bib1,bib2} 
leading to the conclusion that the medium is indeed a plasma of quarks 
and gluons, but it behaves like a ``perfect'' fluid rather than an ideal 
gas~\cite{bib3}. Due to the 
limited sensitivity of the used probes, the conclusions are qualitative 
rather than quantitative. More details on the jet modification in the medium 
are needed and important aspects of the jet-quenching theory are as yet 
untested. 

Heavy quarks (charm, bottom) are an optimal tool to study the medium 
properties since they are primarily produced in the early stage of the 
collision~\cite{bib3} and, therefore, probe the complete space-time evolution 
of the medium. Due to their large mass ($m >$ 1 GeV/c$^2$) their interaction 
processes can be calculated in pQCD. Heavy-quark hadrons live much longer 
($c\tau = 100-200~\mu$m and $400-500~\mu$m for charm and bottom, 
respectively) than the created medium and, therefore, decay outside the 
medium. Theoretical models predicted that heavy quarks should experience 
smaller energy loss than light quarks when propagating through the medium 
due to the mass-dependent suppression (called dead-cone effect)~\cite{bib4}. 

RHIC measurements in heavy-ion collisions, however, have shown~\cite{bib5} 
that the electron yield from semi-leptonic heavy-quark decays is strongly 
suppressed relative to properly scaled p+p collisions, usually quantified 
in the nuclear modification factor. 
This factor shows a similar amount of suppression as observed for 
light-quark hadrons. Energy loss models with reasonable input 
parameters~\cite{bib6,bib7} do not explain the observed suppression. 
It has been shown that collisional energy loss for heavy quarks~\cite{bib8,bib9}
and collisional dissociation of heavy mesons in the medium~\cite{bib9b}
may be significant for heavy-ion collisions. Only theoretical models, 
which include energy loss from charm only, describe the present data 
reasonably well~\cite{bib7}. Since these measurements are sensitive to 
the sum of charm and bottom contributions, there is an urgent need to 
disentangle the relative contributions experimentally.

Recent results on measurements of the azimuthal angular correlations 
between electrons (from heavy-flavor decays) associated with charged 
hadrons have shown that the relative bottom contribution (B/(B+C)) to 
the non-photonic electron spectrum is about 40$\%$ at a transverse 
momentum of $\pt$ = 5 GeV/c~\cite{bib10}. 
The measured $\pt$ dependence of the relative bottom contribution can 
be used to verify the input parameters for most of the energy loss models. 

In this paper, first results on a different approach are presented which 
allow disentangling the charm and bottom contributions to the non-photonic 
electrons using azimuthal correlations of electrons and D$^0$ mesons.

%-----------------------------------------------------------------------------------------
\section{Data analysis}\label{ana}  
The analysis is based on p+p collisions at $\snn$~=~200 GeV 
measured by the STAR experiment at RHIC. Particle identification 
via the specific energy loss (resolution $\sigma_{dEdx}/dEdx = 8\%$) and 
tracking over a large kinematical range with very good momentum 
resolution is performed by the Time Projection 
Chamber (TPC)~\cite{bib11}. The TPC is located inside a solenoidal 
magnet with a maximum field of 0.5T and has an acceptance of 
$|\eta| < 1.4$ and full azimuthal coverage. The STAR detector utilizes 
a barrel-electromagnetic calorimeter (BEMC)~\cite{bib12} as a leading 
particle (electrons, photons) trigger to study high-$\pt$ particle 
production. The BEMC is a lead-scintillator sampling calorimeter with 
an energy resolution of $\delta E/E \approx16\%/\sqrt{E}$. 
The calorimeter, situated behind the TPC, covers an acceptance 
of $|\eta| < 1$ and full azimuth. During Run VI data taking the full 
BEMC was installed and $\approx97\%$ operational.
To enhance the high $\pt$ range, a high-tower trigger was used with 
an energy threshold of 5.4~GeV for the highest energy in a BEMC cell. 
The high-tower trigger efficiency is nearly 100$\%$.\\

%
% giftoppm  file.gif | pnmtops  -rle -dpi 300 -scale 0.25 > file.eps
%

\begin{figure}[htb]
%\vspace*{-.8cm}
\begin{center}
%\includegraphics[width=0.9\textwidth]{fig10.eps}
%\hspace*{2.cm} 
\includegraphics[height=8cm]{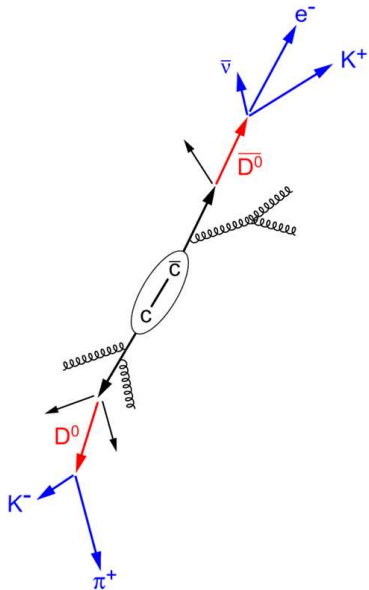}
%\vspace*{-1cm}
\caption[]{\label{fig:1} Schematic view of the fragmentation of $c{\bar c}$ pair.}
\end{center}
%\label{fig:1}
\end{figure}

The integrated luminosity for p+p collisions during Run VI was 
9 pb$^{-1}$ of those 1.2 million events were used after event 
quality cuts (high-tower trigger and z-coordinate of the collision 
vertex (beam axis) within 30~cm of the TPC centre). The tight 
vertex cut is used to minimize the amount of material within 
the detector volume causing photon conversions. \\

Due to the large mass, heavy quarks can only be created in 
pairs such that each quark-antiquark pair is correlated in relative 
azimuth due to momentum conservation. These events are 
characterized by two back-to-back orientated sprays of 
particles leading to a two-peak structure in the azimuthal angular 
correlation distribution. These correlations survive the 
fragmentation process to a large extent in p+p interactions.

A new analysis method is used to identify events with heavy-quark 
production using the back-to-back decay topology of heavy-quark 
jets. As an example, Figure~\ref{fig:1} shows a schematic view 
of the fragmentation of a $c{\bar c}$ pair. The first charm 
particle is identified by 
the subleading electron (trigger side) and the balancing 
charm quark is found by the open charm meson (probe side). 
The branching fraction for a semi-leptonic charm or bottom 
decay ($c, b \rightarrow l + X$) is $\approx10\%$. 
Approximately 54$\%$ of 
the charm quarks decay into D$^0$ mesons. While triggering on 
the subleading electron, the second charm particle can be 
used to identify the underlying production mechanisms since, 
e.g., charm-quark pairs from flavor creation (LO) are 
oriented back-to-back in the azimuthal angular correlation 
distribution ($\dphi = \pi$) whereas they are collinear 
($\dphi = 0$) for gluon splitting (NLO) processes. An 
additional charm contribution on the near-side is expected 
from bottom decays.

In the following sections, the identification of the 
non-photonic electron trigger and the reconstruction of 
the D$^0$ mesons are described in detail.

%-----------------------------------------------------------------------------------------
\subsection{Electron identification}\label{eleid}
The electron identification is performed by combining 
the information from the TPC (track momentum and 
ionization energy loss (dE/dx)) and the BEMC 
(cell energy). Candidate tracks are selected having a 
pseudo-rapidity of $|\eta| < 1$. Due to the momentum 
resolution of the calorimeter, only particles with a 
transverse momentum of $\pt > 1.5$ GeV/c can be 
measured. 

A shower maximum detector, located approximately 
at a depth of 5 radiation length inside the calorimeter 
modules, measures the profile of an electromagnetic 
shower and the position of the shower maximum with 
high resolution ($\deta, \dphi$) = (0.007, 0.007). 
In contrast to hadrons, electrons deposit most of their 
energy in the BEMC cells. A cut on the shower profile 
size combined with a requirement on the ratio of 
momentum-to-cell energy, $0 < p/E < 2$, reject a 
large amount of hadrons. The final electron sample 
is selected by applying a momentum dependent cut 
on the ionization energy loss (about 
$3.5 < dE/dx < 5.0$~keV/cm). More details on the 
electron reconstruction can be found in 
Ref.~\cite{bib5}.

The resulting hadron suppression factor is 10$^5$ 
at $\pt$ = 2 GeV/c and 10$^2$ at $\pt$ = 7 GeV/c. The 
electron purity is $\approx100\%$ at $\pt$ = 5 GeV/c 
and decreases to about 70$\%$ at $\pt$ = 12 GeV/c.

\begin{figure}[htb]
%\vspace*{-.8cm}
%\begin{center}
\subfigure{\includegraphics[width=0.5\textwidth]{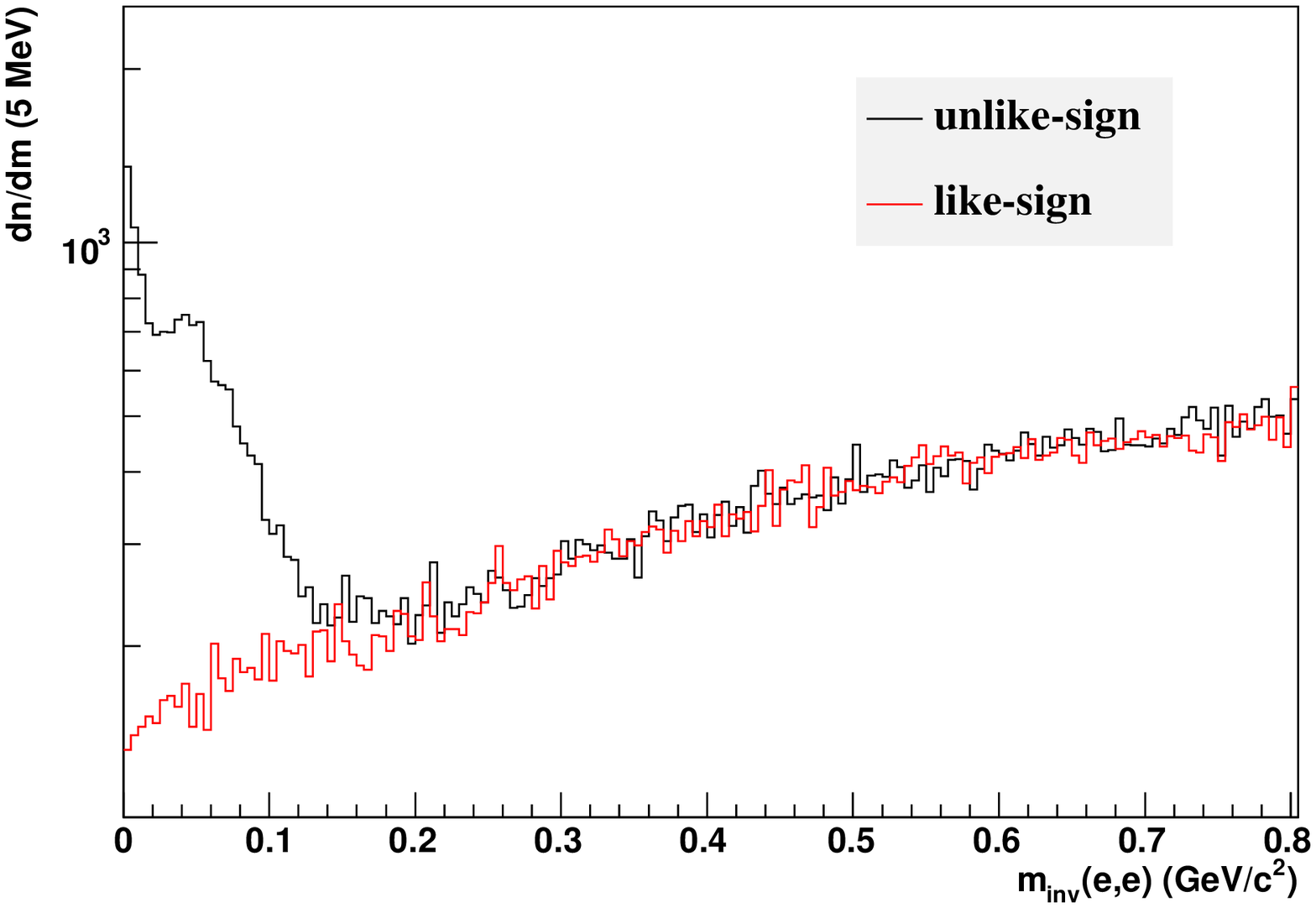}}
\subfigure{\includegraphics[width=0.5\textwidth]{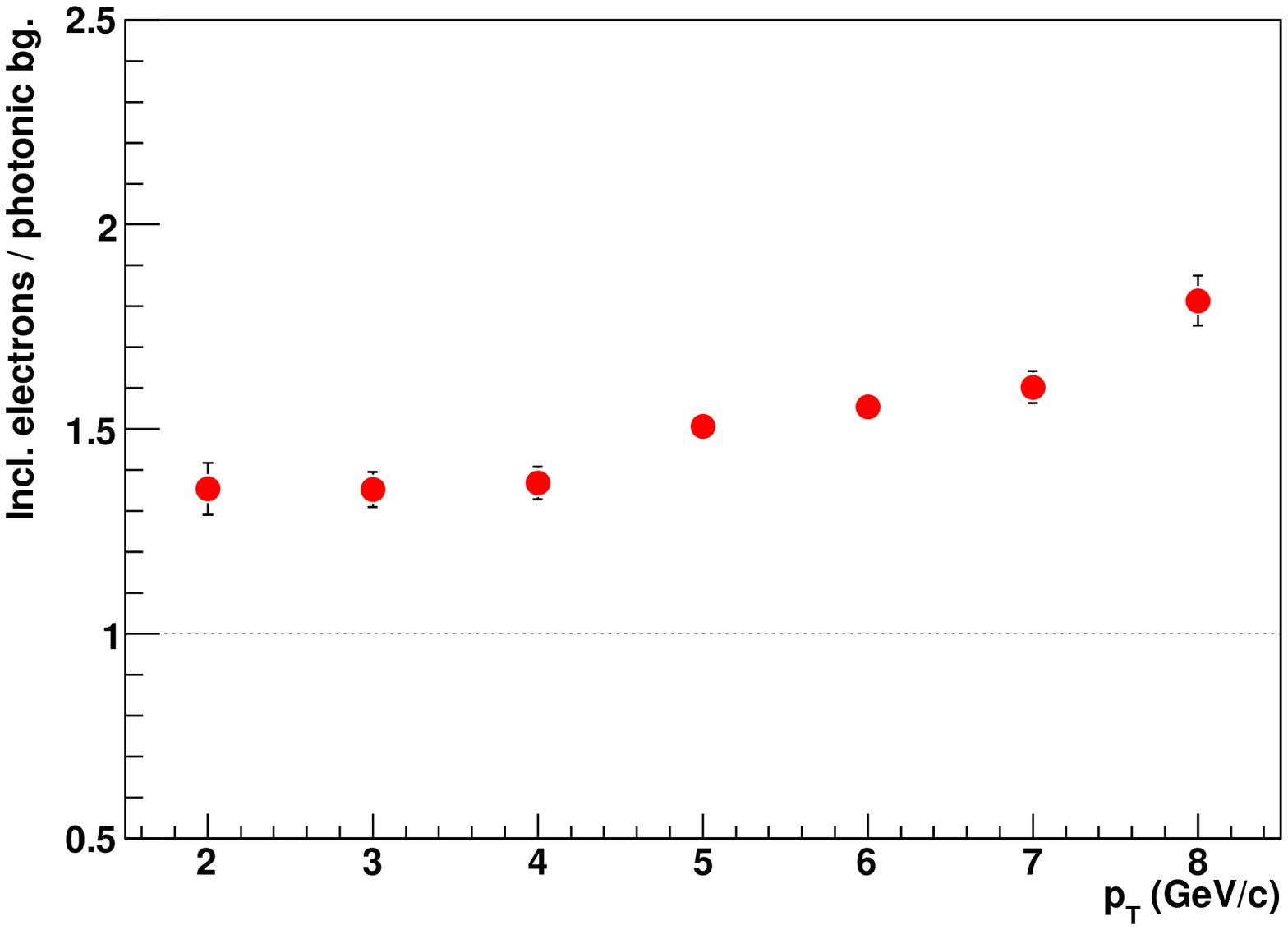}}
\vspace*{-1.cm}
\caption[]{\label{fig:2}
Left panel: Invariant mass distribution of e$^+$e$^-$ 
pairs (black histogram) and the combinatorial background using 
like-sign electron pairs (red histogram). Right panel: Ratio of 
inclusive electrons to the photonic background as a function 
of electron trigger $\pt$. Statistical errors are shown only.}
%\end{center}
%\label{fig:2}
\end{figure}

Most of the electrons are originating from other sources 
than heavy-flavor decays. Photon conversions 
($\gamma \rightarrow e^+e^-$) in the detector material 
between the interaction point and the TPC and neutral 
pion and $\eta$ Dalitz decays ($\pi^0, \eta \rightarrow e^+e^- \gamma$) 
represent the dominant source of photonic electrons. 
Photonic electron contributions from other decays, like 
$\rho$, $\phi$ and Ke3, are small and can be neglected. 
Electrons from the photonic background can be determined 
by calculating the invariant mass of electron pairs. Here, 
each electron candidate is combined with tracks within the 
TPC acceptance which are passing loose cuts on the ionization 
energy loss to preselect electron candidates.

The resulting invariant mass distribution of unlike-sign 
electron pairs is illustrated in the left panel of Figure~\ref{fig:2}. 
The fraction of random pairs with a non-photonic 
electron can be estimated using like-sign pair combinations. 
The peak at zero invariant mass arises from conversions 
and the tail at low invariant mass is due to Dalitz decays~\cite{bib13}. 
Electrons with an invariant mass of $m < 150$~MeV/c$^2$ 
are disregarded. The photonic background finding efficiency, 
using this method, is $\approx70\%$.

The right panel of Figure~\ref{fig:2} shows the $\pt$ 
dependence of the ratio of inclusive electrons to the 
photonic background. A significant excess of non-photonic 
electrons is observed which increases with increasing $\pt$. 
A total of $\approx$6k non-photonic electrons, originating 
mostly from heavy-flavor decays, are obtained for the further 
analysis.

%-----------------------------------------------------------------------------------------
\subsection{D$^0$ reconstruction}\label{D0}
The associated D$^0$ mesons are reconstructed in the hadronic 
decay channel D$^0 \rightarrow$ K$^- \pi^+$ (branching fraction 
3.84 $\%$) 
by calculating the invariant mass of all oppositely charged 
TPC track combinations (unlike-sign pairs) in the same event. 
Up to now, the precise D$^0$ decay topology can not be 
resolved due to insufficient tracking resolution close to the 
collision vertex. A $dE/dx$ cut of $\pm 3\sigma$ around the 
Kaon 
band is applied on the negative tracks. The resulting track 
sample is called Kaon candidates. Due to the high abundance 
of pions in the collisions, one usually has to handle a large 
combinatorial background of random pair combinations~\cite{bib14}. 
Therefore, in this analysis, events with a non-photonic electron 
trigger only are used for the D$^0$ reconstruction. Furthermore, 
the Kaon candidates have to have the same sign as the 
non-photonic electrons. 
%In the following, we imply electron$-$D$^0$ and positron$-$anti-D$^0$ 
%pairs when using e$-$D$^0$ unless otherwise specified.

\begin{figure}[htb]
%\vspace*{-.8cm}
\begin{center}
\includegraphics[width=0.9\textwidth]{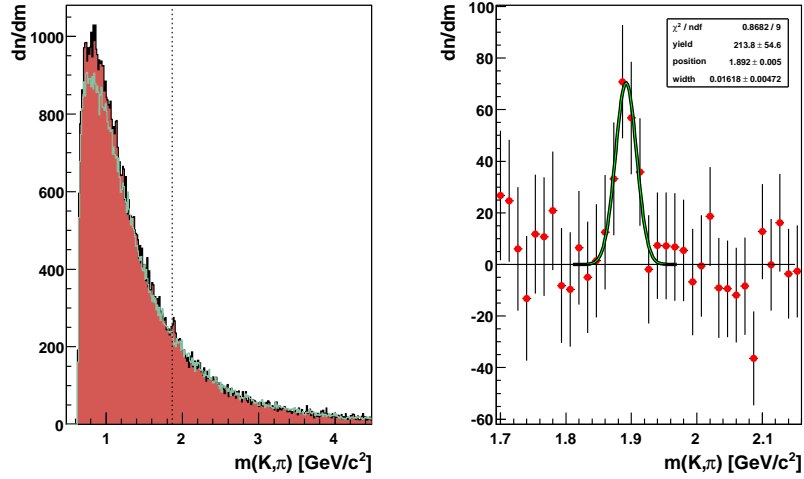}
\vspace*{-.5cm}
\caption[]{\label{fig:3}
Left panel: Invariant mass distribution of (K, $\pi$) pairs 
requiring a non-photonic electron trigger in the event (red histogram) 
and the combinatorial background using like-sign (K, $\pi$) pairs (green 
histogram). Right panel: Background subtracted invariant mass distribution.
The solid black line is a Gaussian fit to the data around the peak region.}
\end{center}
%\label{fig:3}
\end{figure}

The resulting invariant mass distribution of (K, $\pi$) pairs 
shows a clear D$^0$ peak (Figure~\ref{fig:3}, left panel). 
The combinatorial background of random pairs is evaluated by 
combining all like-sign charged tracks in the same event. 
The discrepancy of the shape between the invariant mass 
and the combinatorial background distribution at lower 
masses is due to jet particle correlations which are not 
included in the background evaluation. 
The invariant mass distribution without a non-photonic electron 
trigger does not have a signal for the used track quality cuts. 
The non-photonic electron trigger allows suppressing the combinatorial 
background significantly (by a factor of $\approx100$ compared to 
earlier results~\cite{bib14}) yielding a signal-to-background ratio 
of about 1/7. A signal significance of approximately 4 is obtained. 

In the right panel of Figure~\ref{fig:3}, the background subtracted 
invariant mass distribution is illustrated. The peak position and 
width are obtained using a Gaussian fit to the data. The measured 
peak position, $m = 1.892 \pm$ 0.005 GeV/c$^2$, is slightly higher 
than the PDG value of 1.864 GeV/c$^2$ which can be explained by 
the finite momentum resolution of the TPC. The width of the signal, 
$\sigma_m = 16 \pm$ 5 MeV/c$^2$, 
is found to be similar to earlier results and expectations from 
MC simulations~\cite{bib14}. Within statistical uncertainties, 
the D$^0$ and $\overline{{\rm D}^0}$ yields are equal.

%-----------------------------------------------------------------------------------------
\subsection{Electron$-$D$^0$ meson azimuthal correlation}\label{corr}
\begin{figure}[htb]
%\vspace*{-.8cm}
\begin{center}
\includegraphics[width=0.9\textwidth]{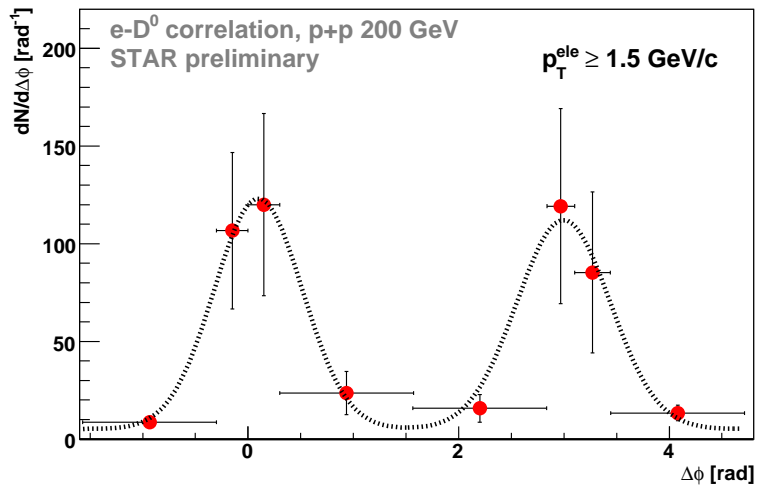}
%\insertplot{analysis_cuts.eps}
\vspace*{-0.5cm}
\caption[]{\label{fig:4}
Azimuthal angular correlation distribution of non-photonic 
electrons and D$^0$ mesons in p+p collisions at $\snn$~=~200 GeV
(trigger $\pt \geq 1.5$ GeV/c). Statistical errors are shown only. 
To guide the eye, the dotted line illustrates a two-Gaussian plus 
constant fit to the data.}
\end{center}
%\label{fig:4}
\end{figure}

The azimuthal angular ($\dphi$) correlation is calculated between 
the transverse momentum of the non-photonic electron triggers 
and the associated charged hadron-pairs. The (K, $\pi$) 
invariant mass distribution is calculated for different $\dphi$ 
bins and the yield of the associated D$^0$ mesons is extracted as 
the area underneath a Gaussian fit to the signal. 
Figure~\ref{fig:4} shows the e$-$D$^0$ azimuthal correlation distribution. 
In the following, we imply electron$-$D$^0$ and positron$-$anti-D$^0$ 
pairs when using e$-$D$^0$ unless otherwise specified.

%-----------------------------------------------------------------------------------------
\section{Results}\label{results}  
The e$-$D$^0$ azimuthal correlation distribution exhibits a near- 
and away-side correlation peak with similar yields. The data shows 
indications for prompt charm meson pair production leading to 
the back-to-back correlation peak ($\dphi = \pi$). The near-side 
correlation peak can be interpreted by contributions from gluon 
splitting and bottom decays. The CDF collaboration obtained a 
similar correlation pattern for D$^0$(D$^+$)$-$D$^{*-}$ pairs in 
p+$\overline{\rm p}$ collisions at $\snn$~=~1.96 TeV at the Tevatron 
Collider at Fermilab~\cite{bib15}. Their measurements have shown that 
gluon splitting ($\dphi = 0$) is as important as flavor creation 
processes ($\dphi = \pi$). 
Related studies have yet to be performed using these data and event 
generators such as Pythia and Herwig to obtain the relative contributions 
at RHIC energies.
%Calculations from event generators, e.g., PYTHIA and Herwig, have to 
%show what the relative contributions are at RHIC energies.

%--------------------------------------------------------------------------------------- 
\section{Conclusions}\label{concl}
This work presents the first heavy-flavor correlation measurement at RHIC. 
The particular advantage of the presented correlation method, in contrast to 
conventional open charm measurements, is the possibility to trigger on 
collisions with heavy-quark production using the decay electrons. 

The azimuthal correlation distribution between non-photonic electrons associated 
with open charm mesons exhibits a two-peak structure. The away-side correlation peak 
can be explained by prompt charm meson pair production whereas the near-side peak 
represents contributions from gluon splitting and bottom decays. 
The results are in qualitative agreement with recent CDF measurements. 
Dedicated event generators, like Pythia or Herwig, are needed to disentangle the 
charm and bottom contribution to the azimuthal correlation distribution. 

The presented, new correlation technique has the potential for detailed energy loss 
measurement of open charm mesons in heavy-ion collisions in the future.

%--------------------------------------------------------------------------------------- 
\section*{Acknowledgments}
The author is grateful for the support by the Nederlands Organisation for Scientific 
Research (NWO).

%--------------------------------------------------------------------------------------- 

\vfill\eject
\end{document}